\documentclass[conference]{IEEEtran}
\IEEEoverridecommandlockouts
\usepackage{cite}

\usepackage{amsmath,amssymb,amsfonts}
\usepackage{algorithmic}
\usepackage{graphicx}
\usepackage{textcomp}
\usepackage{xcolor}
\usepackage{bm}
\usepackage[caption=false]{subfig}
\usepackage{xurl}

\def\BibTeX{{\rm B\kern-.05em{\sc i\kern-.025em b}\kern-.08em
    T\kern-.1667em\lower.7ex\hbox{E}\kern-.125emX}}

\begin{document}

\title{Learning-Based Interface for Semantic Communication with Bit Importance Awareness
\thanks{This work was supported by the National Natural Science Foundation of China under Grant No. 62231022. Corresponding author: Wenyi Zhang.

The code is available at https://github.com/perseconds/InterfaceSC}
}

\author{
\IEEEauthorblockN{Wenzheng Kong and Wenyi Zhang}
\IEEEauthorblockA{\textit{ Department of Electronic Engineering and Information Science} \\
\textit{University of Science and Technology of China}\\
Hefei 230027, China\\
wenzhengk@mail.ustc.edu.cn, wenyizha@ustc.edu.cn}
}

\maketitle

\begin{abstract}
Joint source-channel coding (JSCC) is an effective approach for semantic communication. However, current JSCC methods are difficult to integrate with existing communication network architectures, where application and network providers are typically different entities. Recently, a novel paradigm termed Split DeepJSCC has been under consideration to address this challenge. Split DeepJSCC employs a bit-level interface that enables separate design of source and channel codes, ensuring compatibility with existing communication networks while preserving the advantages of JSCC in terms of semantic fidelity and channel adaptability. In this paper, we propose a learning-based interface design by treating its parameters as trainable, achieving improved end-to-end performance compared to Split DeepJSCC. In particular, the interface enables specification of bit-level importance at the output of the source code. Furthermore, we propose an Importance-Aware Net that utilizes the interface-derived bit importance information, enabling dynamical adaptation to diverse channel bandwidth ratios and time-varying channel conditions. Experimental results show that our method improves performance in wireless image transmission tasks. This work provides a potential solution for realizing semantic communications in existing wireless networks.
\end{abstract}

\begin{IEEEkeywords}
Deep learning, digital interface, JSCC, semantic communication, wireless image transmission
\end{IEEEkeywords}

\section{Introduction}
 As a key technology in sixth-generation (6G) wireless networks, semantic communication has attracted extensive attention in recent years. The concept was initially discussed by Weaver \cite{weaver1963mathematical}, who pointed out the importance of preserving the feature of information in contrast to conventional communication systems that emphasize accurate transmission of symbols. Driven by recent advances in deep learning (DL) technologies, DL-based Joint Source-Channel Coding (JSCC) has emerged as a promising approach for enabling semantic communication. Existing studies have shown that DL-based JSCC can enhance information transmission efficiency and achieve improved end-to-end performance in various modalities, such as text \cite{xie2021deep}, image \cite{bourtsoulatze2019deep}, video \cite{tung2022deepwive} and speech \cite{weng2021semantic}.

Despite the promising end-to-end performance of DL-based JSCC, its practical deployment in existing communication networks still faces significant challenges. Existing communication networks employ layered architectures which enable independent development and optimization of each layer. Transmitted information is initially processed at the application layer of the source node, resulting in data packets. Then the data packets pass through backbone networks with stable links before arriving at the final noisy wireless hop. Such a system architecture necessitates separate design of source coding (by application providers) and channel coding (by network providers), and the two coding processes are often implemented at geographically distributed nodes. Additionally, some JSCC schemes requiring real-time channel state information (CSI) are difficult to implement, as CSI must be sent back through multiple hops to the source node, which is hindered by intermediate nodes (e.g., base stations or access points) lacking access to application’s source.

To address these challenges, DL-driven JSCC naturally calls for decoupling to enable separate design of source and channel codes. Recently, a decoupled JSCC framework named Split DeepJSCC \cite{tung2025multi} has been proposed to overcome limitations of implementing JSCC over today’s communication networks. Split DeepJSCC introduces a multi-level reliability binary interface to enable separate design of DL-driven source and channel codes. In particular, the interface specifies the reliability level of different bit positions at the output of the source code. Split DeepJSCC preserves the advantages of end-to-end JSCC while maintaining compatibility with existing communication networks. However, its binary interface relies on prescribed configurations that cannot utilize the relationship between image reconstruction quality and bit importance. Consequently, Split DeepJSCC still has significant room for improvement, and we propose a learning-based interface design that treats its parameters as trainable, eliminating explicit reliance on prescribed configurations. Furthermore, we introduce an Importance-Aware Net (IAN) that utilizes the bit-level importance information specified by the interface, enabling an adaptive coding strategy in face of different channel bandwidth ratios (CBRs) and time-varying channel quality. Experimental results for wireless image transmission demonstrate that the proposed method outperforms Split DeepJSCC across various image categories, different CBRs, and diverse channel conditions.
\vspace{-0.8em} 
\section{System Model}
\begin{figure}[htbp]
\centerline{\includegraphics[width=0.5\textwidth]{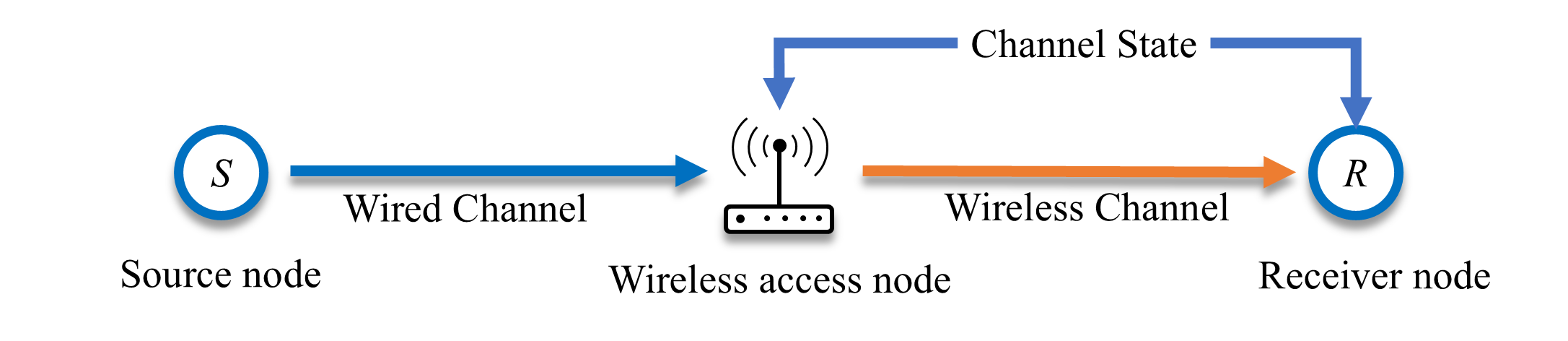}}
\caption{System model.}
\label{fig:systemModel}
\end{figure}

We consider a system model abstracted as a three-node communication network, similar to that in \cite{tung2025multi}, where a source node transmits information to a receiver node through a wireless access node. As shown in Fig.~\ref{fig:systemModel}, the connection between the source and wireless access nodes represents a wired link modeled as an error-free bit pipe, while the path from the wireless access node to the receiver node represents a wireless link distorted by channel fading and noise. The proposed model conforms more closely to practical communication networks: the source node (representing the application layer) performs source coding to eliminate redundancy, converting original information into bits. The wireless access node (modeling lower network layers of an intermediate node) subsequently applies channel coding to enhance error resilience for wireless transmission. At the receiver node, the decoding process reverses encoding operations through channel and source decoding to reconstruct the original information.

\subsection{Overall Architecture}
\begin{figure}[htbp]
\centerline{\includegraphics[width=0.5\textwidth]{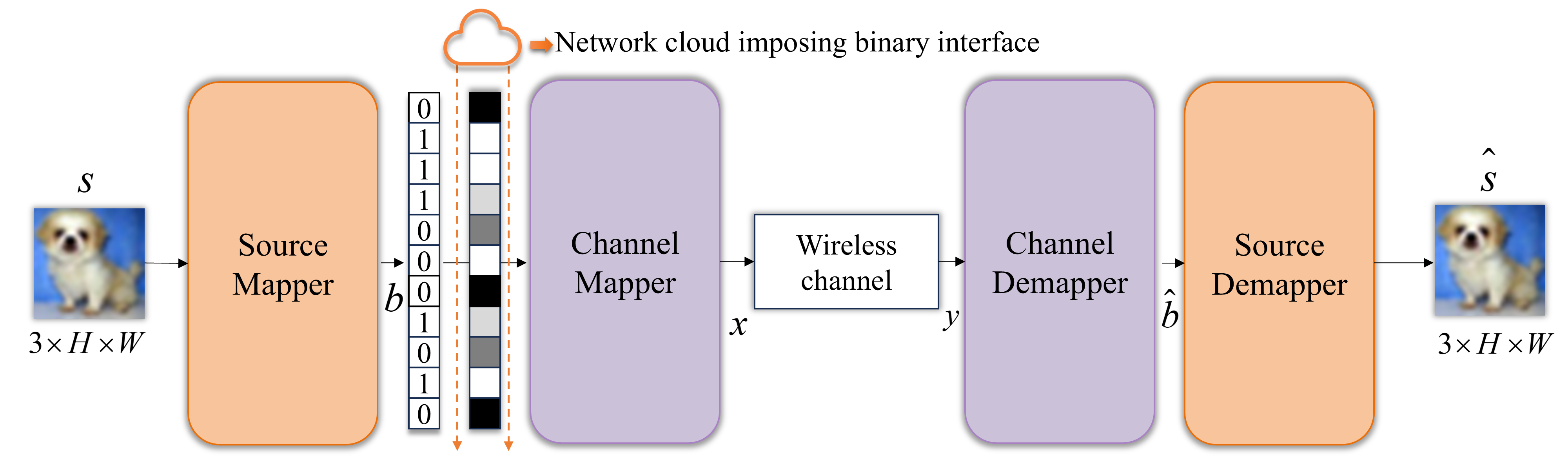}}
\caption{Overall architecture of system model. The network cloud imposes the binary interface to decouple the design of source and channel codes. The interface specifies the parameters of BSCs for the source mapper and demapper, and serves as inputs for the channel mapper and demapper.}
\label{fig:overview}
\end{figure}

The overview of our system model is illustrated in Fig.~\ref{fig:overview}. In our model,
the source and channel mappings are separately designed via a shared agreement between the source and wirelss access nodes. The source and wireless access nodes separately train their coding strategies without joint design. Specifically, this shared agreement, referred to as a binary interface, quantifies the bit-level importance through the bit-flipping probabilities of an array of binary symmetric channels (BSCs). The encoding at the source and wireless access nodes is performed by the source and channel mappers, respectively. At the receiver node, the decoding process connects the channel demapper and the source demapper. Deep neural networks are employed to parameterize the source/channel mapper/demapper.

Let $\bm{s} \in \mathbb{R}^{H \times W \times C}$ denote the image data and let $\bm{\theta}$ represent the parameters of the source mapper. The source mapper is employed to transform the image $\bm{s}$ into a bit sequence $\bm{b}$, ensuring compatibility with the interface and existing communication networks. Since this mapping process is non-differentiable, we adopt a variational learning approach that transforms the source mapper into a generative process. During training, the source mapper is treated as a probabilistic encoder $p_{\bm{\theta}}(\bm{b'}|\bm{s})$, where each output bit follows an independent Bernoulli distribution defined by the function $f_{\bm{\theta}}$ that generates the probability of the corresponding bit being 1. 
After training, during transmission, the probabilistic encoder is replaced by its deterministic version through element-wise rounding operations to yield binary encoder, i.e.,
\begin{equation}
    \bm{b} = \arg\max_{\bm{b'}} p_{\theta}(\bm{b'}|\bm{s}) = \left\lfloor f_{\bm{\theta}}(\bm{s}) \right\rceil \in \{0, 1\}^M,
    \label{sourcemapper}
\end{equation}
where $\left\lfloor \cdot \right\rceil$ denotes the element-wise rounding operation and $f_{\bm{\theta}}$ denotes the source mapper function. Subsequently, the channel mapper with parameters $\bm{\eta}$ maps the bit sequence $\bm{b}$ into $L$ complex-valued channel input symbols $\bm{x}$, denoted as:
\begin{equation}
    \bm{x} = g_{\bm{\eta}}(\bm{b}) \in \mathbb{R}^L,
\end{equation}
where $g_{\bm{\eta}}$ is the channel mapper function.
Furthermore, the CBR is defined as: 
\begin{equation}
    r=L/(C \times H \times W).
\end{equation}

We normalize the power of $\bm{x}$ to ensure the channel inputs to satisfy an average power constraint. In this paper, we model the wireless channel between the wireless access node and the receiver node as a flat fading channel. Accordingly, the received signal $\bm{y}$  at the receiver node is expressed as:
\begin{equation}
    \bm{y} = \bm{h} \odot \bm{x} + \bm{n},
\end{equation}
where $\odot$ is the element-wise product, $\bm{h} \in \mathbb{C}$ is the complex-valued channel coefficient, and $\bm{n} \sim \mathcal{CN}(0, \sigma_n^{2}\bm{I_k})$ denotes the additive white Gaussian noise (AWGN). We assume that the $\bm{h}$ is accurately estimated at the receiver.

At the receiver node, the received signal $\bm{y}$ is first processed by the channel demapper to produce a bit sequence $\bm{\hat{b}}$. Let $\bm{\gamma}$ represent the parameters of the channel demapper. Consistent with the source mapper, the channel demapper is modeled as probabilistic decoder $p_{\bm{\gamma}}(\bm{\hat b'}|\bm{y})$ during training, while converted to a deterministic binary decoder during transmission, i.e.,
\begin{equation}
    \bm{\hat{b}} = \arg\max_{\bm{\hat{b'}}} p_{\bm{\gamma}}(\bm{\hat b'}|\bm{y}) = \left\lfloor g_{\bm{\gamma}}(\bm{y}) \right\rceil \in \{0, 1\}^M,
\end{equation}
where $g_{\bm{\gamma}}$ denotes the channel demapper function, producing the probability of the corresponding bit being 1. Then the recovered bit sequence $\bm{\hat{b}}$ passes through the source demapper to reconstruct the original image data $\bm{s}$, denoted by $\bm{\hat{s}}$, as follows:
\begin{equation}
    \bm{\hat s} = f_{\bm{\phi}}(\bm{\hat{b}}) \in \mathbb{R}^{H \times W \times C},
\end{equation}
where $f_{\bm{\phi}}$ denotes the source demapper function with parameters $\bm{\phi}$.
\begin{figure*}[t]
    \centering
    \subfloat[Training Source Mapper and Demapper via Interface]{
        \includegraphics[width=0.48\linewidth]{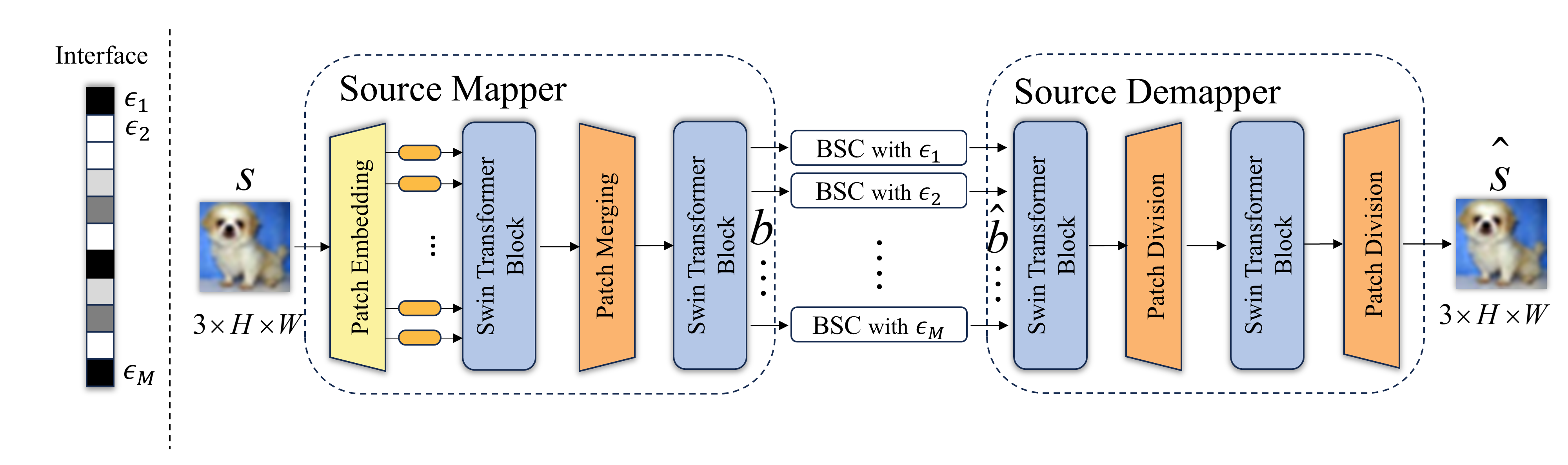}
        \label{fig:TrainSource}
    }
    \hfill
    \subfloat[Training Channel Mapper and Demapper]{
        \includegraphics[width=0.48\linewidth]{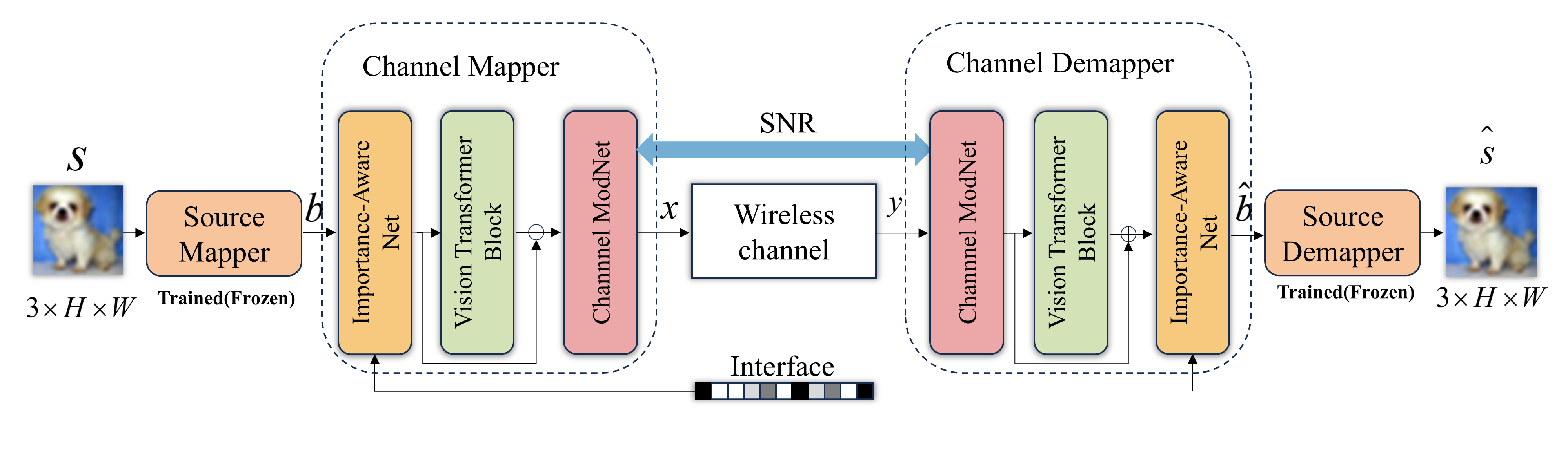}
        \label{fig:TrainChannel}
    }
    \caption{Illustration of model training.}
    \label{fig:pair}
\end{figure*}

To evaluate the quality of the received image, we employ the peak signal-to-noise ratio (PSNR), defined as:
\begin{equation}
\mathrm{PSNR}(\bm{s},\bm{\hat{s}}) = 10\log_{10}\left(\frac{\mathrm{MAX}^2}{\mathrm{MSE}(\bm{s},\bm{\hat{s}})}\right) \, \text{dB},
\end{equation}
where $\mathrm{MAX}$ represents the maximum possible value of each element in $\bm{s}$ (e.g., 255 for an 8-bit RGB pixel), and the mean-squared error (MSE) is calculated as $\mathrm{MSE}(\bm{s},\bm{\hat{s}}) = \|\bm{s} - \bm{\hat{s}}\|_2^2$. Notably, unlike traditional separate source-channel coding (SSCC), our approach does not prioritize accurate reconstruction of the bit sequence $\bm{\hat{b}}$, and instead focuses on image reconstruction, enabling semantic-aware transmission. Therefore, our approach should still be classified into the paradigm of JSCC, just as Split DeepJSCC \cite{tung2025multi}.

The proposed system employs a two-stage training strategy due to the separate encoding at the source and wirele access nodes. In the first stage, the source mapper and demapper are trained through the BSCs, whose bit-flipping probabilities are specified by the interface. In particularly, the bit-flipping probabilities are treated as trainable parameters and are concurrently optimized during training. Once trained, the parameters of the source mapper and demapper are frozen and the interface parameters are fixed. In the second stage, channel mapper and demapper are trained by utilizing interface-derived bit importance information to minimize the loss observed from the source demapper over wireless channel. We emphasize that, in the first stage, the interface is used to specify the bit-flipping probabilities of the BSCs, so as to achieve effective training of the source mapper and demapper, but the BSCs are subsequently removed after training is completed; in the second stage, values of parameters of the trained interface are then fed into the channel mapper and demapper to provide the bit importance information.

\subsection{Training Source Mapper and Demapper}
Inspired by the recent advances of vision Transformer \cite{dosovitskiy2020image} in the field of computer vision, we adopt the Swin Transformer \cite{liu2021swin} as the backbone architecture for both the source mapper and demapper. The detailed structure is illustrated in Fig.~\ref{fig:TrainSource}. 

In our system, we train the source mapper and demapper with the BSCs, whose bit-flipping probabilities are specified by the interface, as shown in Fig.~\ref{fig:TrainSource}. The BSCs serve as a model to characterize the stochastic transformation from $\bm{b}$ to $\bm{\hat{b}}$, aiming to emulate the combined effects of the channel mapper, wireless channel and channel demapper. We construct an array of $M$ BSCs with bit-flipping probabilities $\{\epsilon_n\}_{n=1}^M$, where $M$ equals the length of the bitstream $\bm{b}$. In this model, the conditional probability distribution of $\hat{b}_n$ given $b_n$ is reprensented as: 
\begin{equation}
p_{\text{BSC}}(\hat{b}_n | b_n) = 
\begin{cases}
\epsilon_n, & \text{if} \quad  \hat{b}_n \neq b_n,\\
 1-\epsilon_n, & \text{if} \quad \hat{b}_n = b_n.
\end{cases}
\end{equation}

The design of the bit-flipping probabilities $\{\epsilon_n\}_{n=1}^M$ is critical, as they significantly influence image reconstruction quality. However, the relationship between bit-flipping probabilities and reconstructed image quality is mathematically intractable, and we are hence motivated to adopt a learning-based approach for interface parameter optimization, as opposed to the prescribed interface specification in Split DeepJSCC \cite{tung2025multi}. Inspired by \cite{oh2025digital}, we treat bit-flipping probabilities $\{\epsilon_n\}_{n=1}^M$ as trainable parameters. To prevent model collapse into error-free solutions (i.e., $\epsilon_n \to 0$), we impose an regularization constraint as follows:
\begin{equation}
\mathcal{L}_{\text{reg}} = \frac{\lambda}{M} \sum_{n=1}^M (\epsilon_n - 0.5)^2,
\end{equation}
where $\lambda$ governs the strength of regularization. This formulation promotes convergence of $\epsilon_n$ to higher bit-flipping probabilities, endowing source mapper and demapper with a certain extent of distortion resilience.

In order to effectively learn the source mapper and demapper, we maximize the mutual information between the input $\bm{s}$ and the corresponding noisy bit sequence $\bm{\hat{b}}$. Following the strategy in \cite{choi2019neural, park2024joint}, the channel from $\bm{s}$ to $\bm{\hat{b}}$ can be described using the following conditional probability distribution:
\begin{align}
q_{\text{noisy}}(\bm{\hat{b}}|\bm{s}) 
&= \sum\limits_{\bm{b} \in \{0,1\}^M} 
    p_{\bm{\theta}}(\bm{b}|\bm{s}) \,
    p_{\text{BSC}}(\bm{\hat{b}}|\bm{b}) \\
&= \prod_{n=1}^M 
\underbrace{
\sum_{b_n \in \{0,1\}} 
    p_{\bm{\theta}}(b_n|\bm{s}) \, p_{\text{BSC}}(\hat{b}_n|b_n)
}_{p_{\text{noisy}}(\hat{b}_n|\bm{s})}.
\end{align}
where $p_{\text{noisy}}(\hat{b}_n|\bm{s})$ is further calculated as:
\begin{align}
    p_{\text{noisy}}(\hat{b}_n|\bm{s})=&
    [f_{\bm{\theta}}(\bm{s})_n(1 - \epsilon_n) + \big(1- f_{\bm{\theta}}(\bm{s})_n\big) \epsilon_n]^{\hat{b}_n} \nonumber \\
    &[f_{\bm{\theta}}(\bm{s})_n\epsilon_n +  \big(1- f_{\bm{\theta}}(\bm{s})_n\big) (1 - \epsilon_n)]^{1-\hat{b}_n}.
    \label{sample}
\end{align}
During the training process, the source demapper's input $\bm{\hat{b}}$ is sampled according to \eqref{sample}. Then, we formulate the following optimization problem:
\begin{flalign}
\max_{\bm{\theta},\bm{\epsilon}} I(\bm{s};\bm{\hat{b}})= \mathbb{E}_{\bm{s} \sim p_{\text{data}}(\bm{s})} \left[\mathbb{E}_{\bm{\hat{b}}\sim q_{\text{noisy}}(\bm{\hat{b}}|\bm{s})} \left[\log p(\bm{s}|\bm{\hat{b}})\right]\right] + \text{const},
\end{flalign}
where $p(\bm{s}|\bm{\hat{b}})$ is the true posterior distribution. Since this distribution is intractable, we use a variational approximation $p_{\bm{\phi}}(\bm{s}|\bm{\hat{b}})$ by assuming that the source demapper $f_{\bm{\phi}}$ is a stochastic decoder whose output follows a Gaussian distribution with mean $f_{\bm{\phi}}(\bm{\hat{b}})$. Thus maximizing the mutual information is equivalent to minimizing the mean squared error (MSE) between $\bm{s}$ and $\bm{\hat{s}}$, given as:
\begin{equation}
\mathcal{L}_{\text{MSE}} = 
\mathbb{E}_{\bm{s} \sim p_{\text{data}}(\bm{s})} \left[ 
    \mathbb{E}_{\bm{\hat{b}} \sim q_{\text{noisy}}(\bm{\hat{b}}\mid \bm{s})} \left[
    \big\| \bm{s} - f_{\bm{\phi}}(\bm{\hat{b}}) \big\|^2 \right]
\right]. \label{mse}
\end{equation}

Since direct sampling from the distribution $q_{\text{noise}}$ is non-differentiable, we adopt the Straight-Through Estimator (STE) \cite{bengio2013estimating} technique to allow gradient backpropagation through the sampling process. Consequently, the loss function for the first training stage is given as:
\begin{equation}
    \mathcal{L}(\bm{\theta}, \bm{\phi},\bm{\epsilon}) = 
    \mathcal{L}_{\text{MSE}} + \lambda \mathcal{L}_{\text{reg}}.
    \label{reg}
\end{equation}

\subsection{Training Channel Mapper and Demapper}
Once the first training stage has been completed, the source mapper and demapper are frozen. The source mapper is converted to a binary encoder as described in \eqref{sourcemapper} and the interface parameters keep fixed. In the second training stage, we optimize the channel mapper and demapper by utilizing interface-derived bit importance information to minimize the loss observed from the source demapper over wireless channel, as illustrated in Fig.~\ref{fig:TrainChannel}. The Transformer architecture \cite{dosovitskiy2020image} is adopted as the backbone to reduce computational overhead. To adapt to varying channel conditions, we integrate the ChannelModNet proposed in \cite{yang2023witt} for dynamic channel adaptation. Specifically, we introduce an interface-derived module, i.e., the IAN, which guides the channel mapper and demapper to focus on bit positions of higher importance, enhancing end-to-end performance.
\begin{figure}[htbp]
    \centering
    \subfloat[Attention Module]{
        \includegraphics[width=0.9\linewidth]{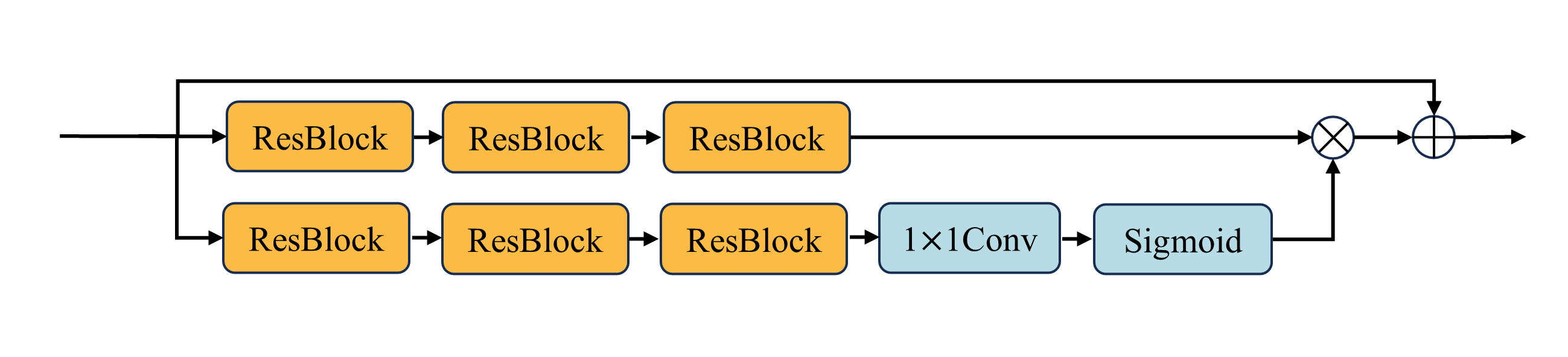}
        \label{fig:attention}
    }

    \subfloat[Importance-Aware Net]{
        \includegraphics[width=0.9\linewidth]{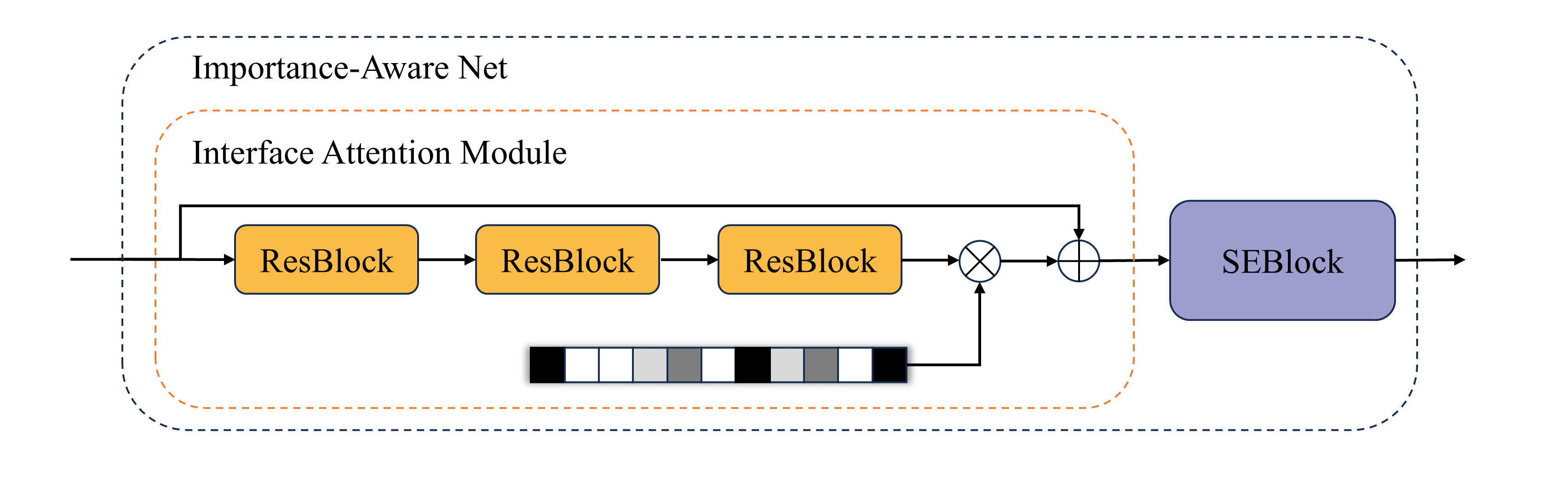}
        \label{fig:importance}
    }
    
    \caption{Structure of the attention module.}
    \label{fig:ImportanceAware}
\end{figure}

The structure of the IAN is depicted in Fig.~\ref{fig:importance}. For the interface attention module, we modify the conventional convolutional attention mechanism (shown in Fig.~\ref{fig:attention}) by directly integrating interface information into latent features. As previously described, the interface is characterized by bit-flipping probabilities, which implicitly reflect the importance of each bit, with more important bits corresponding to lower bit-flipping probabilities. To conform to the attention score mechanism, the bit-flipping probabilities are inverted through the transformation $1 - 2\bm{\epsilon}$, where $\bm{\epsilon}$ denotes the bit-flipping probabilities. Furthermore, a Squeeze-and-Excitation (SE) block \cite{hu2018squeeze} is incorporated to capture image inter-channel relationships. Ablation study in Section \ref{ablation} confirms that the IAN effectively enhances the performance of the channel mapper and demapper.

The channel demapper is modeled as a probabilistic decoder $p_{\bm{\gamma}}(\bm{\hat b}|\bm{y})$ during training. Notably, unlike conventional channel coding systems designed for accurate bit reconstruction, the proposed system prioritizes image recovery through end-to-end quality-aware optimization. This preserves semantic-awareness in the channel mapper and demapper while decoupling them from the source mapper and demapper. Consequently, similar to \eqref{mse}, the loss function for the second training stage is given as:
\begin{align}
\mathcal{L}_{\text{channel}}&= 
   \mathbb{E}_{\bm{s} \sim p_{\text{data}}(\bm{s})} \left[ 
    \mathbb{E}_{\bm{\hat{b}} \sim p_{\bm{\gamma}}(\bm{\hat{b}}\mid \bm{y})} \left[
    \big\| \bm{s} - f_{\bm{\phi}}(\bm{\hat{b}}) \big\|^2 \right] \right].
\end{align}
For non-differentiable processes originating in sampling $\bm{\hat{b}}$ from the $p_{\bm{\gamma}}$ distribution, we still adopt the STE \cite{bengio2013estimating} technique to enable gradient backpropagation.

\section{Experimental Results}
\subsection{Experimental Setup}
\subsubsection{Datasets}
We train and evaluate our proposed  model on four image datasets: CIFAR-10\cite{2009learning}, CIFAR-100\cite{2009learning}, SVHN\cite{netzer2011reading}, and ImageNet32\cite{chrabaszcz2017downsampled}. The CIFAR10 and CIFAR100 datasets both consist of 60,000 color images of size $32 \times 32 \times 3$ pixels, of which 50,000 are used for training and 10,000 for testing. We train the model using the training samples from CIFAR10 and evaluate the performance on the test samples from CIFAR10 and CIFAR100 respectively. The SVHN dataset consists of approximately 630,000 color images of size $32 \times 32 \times 3$ pixels, with 604,000 used for training and 26,000 for testing, capturing street-view house numbers from real-world environments. The ImageNet32 datasets comprises 1.28 million color images of size $32 \times 32 \times 3$ pixels, with 1.28 million for training and 50,000 for testing, encompassing real-world scenes across 1,000 diverse object categories. To accelerate training efficiency, we randomly sample 60,000 images from its original 1.28-million training set for training, while retaining the full 50,000 test images for evaluation. 
\begin{figure*}
    \centering
    \subfloat[CIFAR10]{%
        \includegraphics[width=0.24\linewidth]{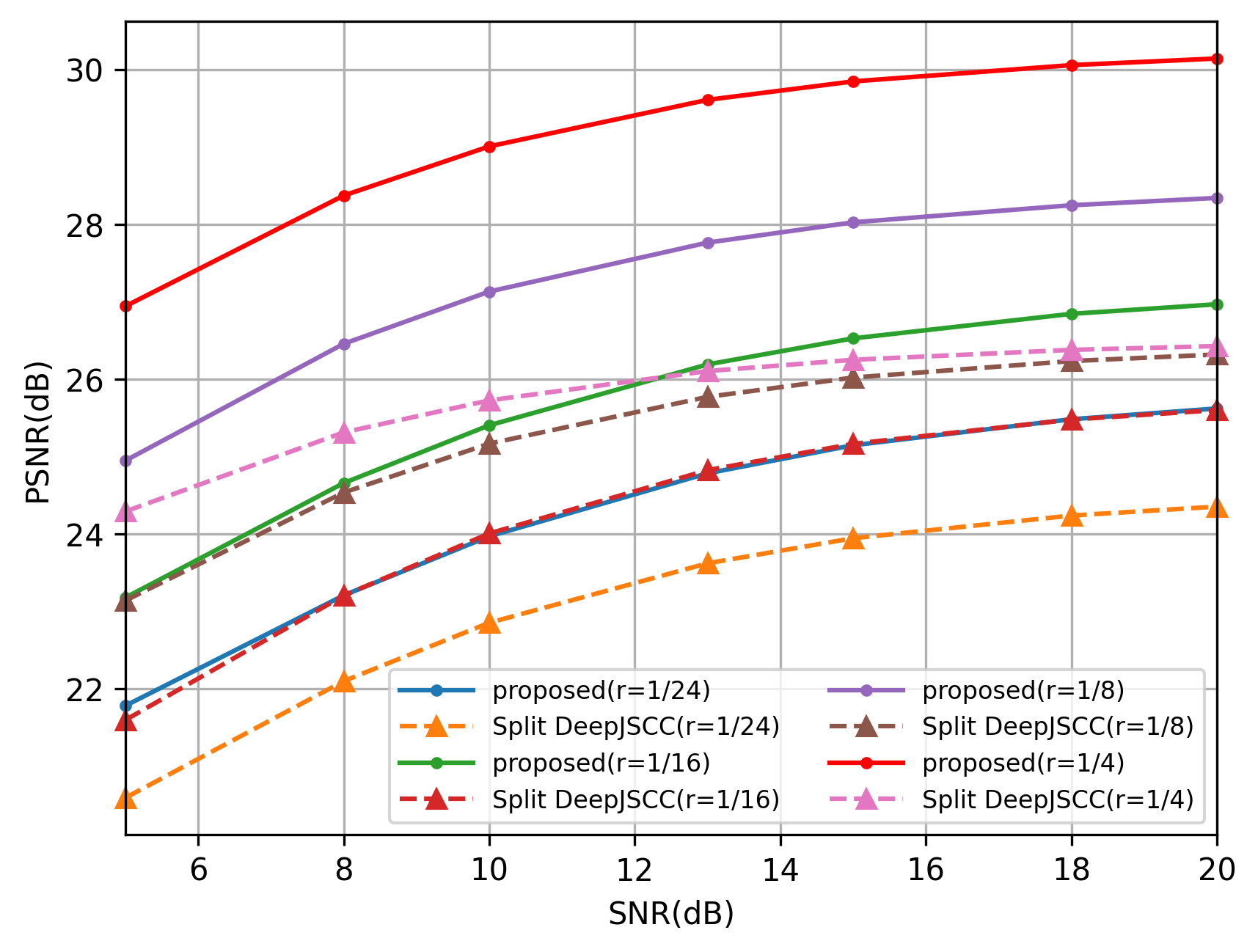}%
        \label{fig:AwgnCIFAR10}%
    }
    \hfil
    \subfloat[CIFAR100]{%
        \includegraphics[width=0.24\linewidth]{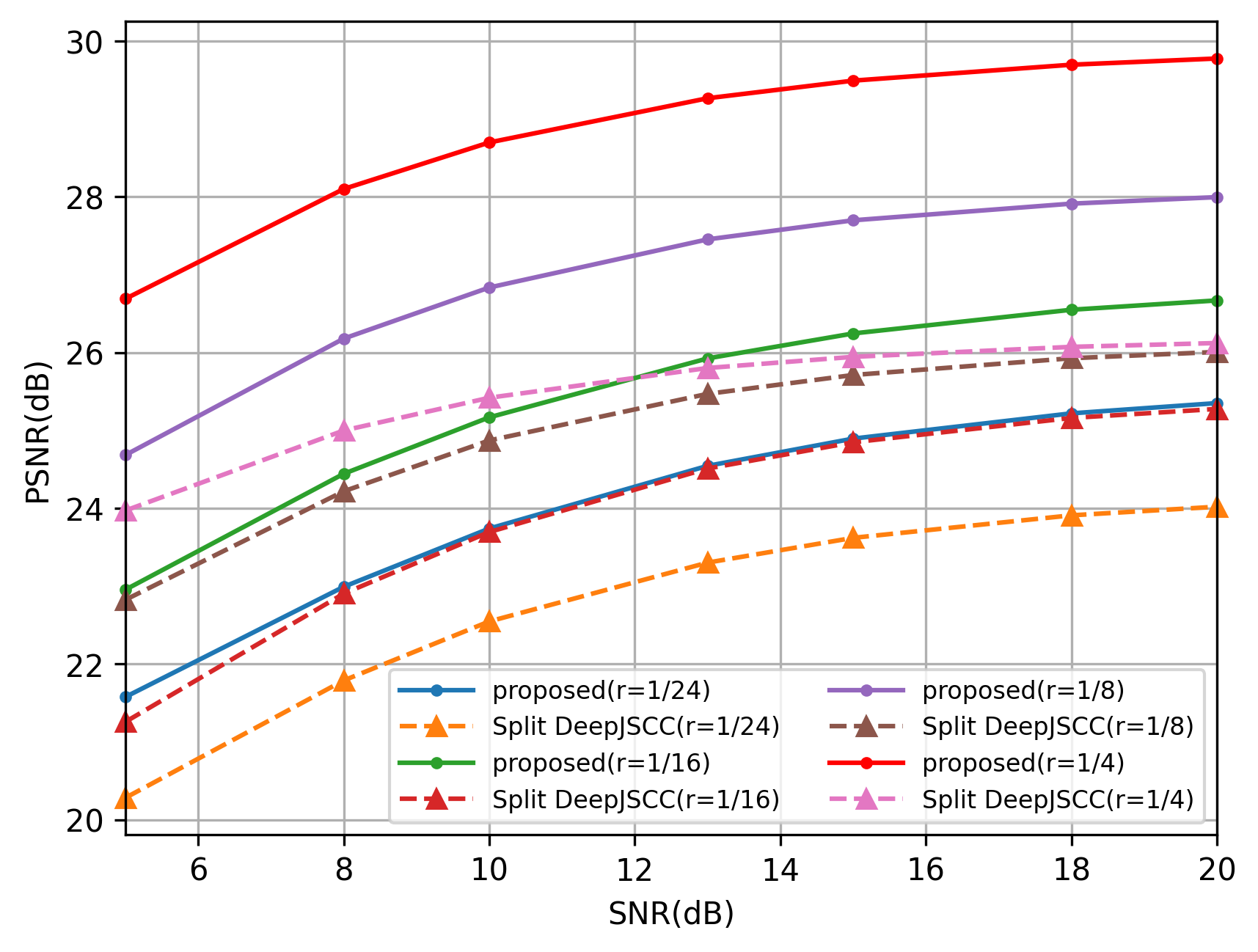}%
        \label{fig:AwgnCIFAR100}%
    }
    \hfil
    \subfloat[SVHN]{%
        \includegraphics[width=0.24\linewidth]{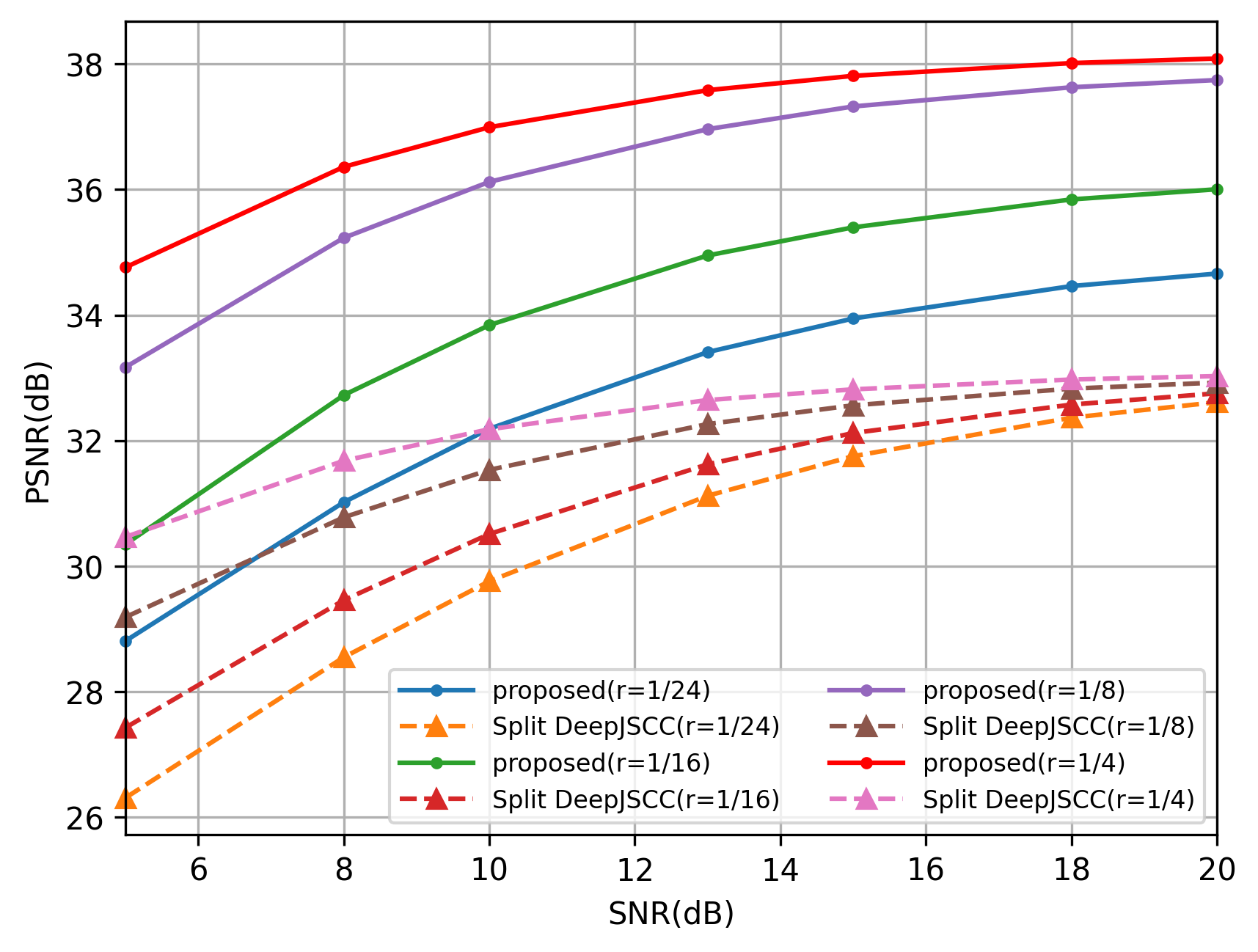}%
        \label{fig:AwgnSVHN}%
    }
    \hfil
    \subfloat[ImageNet32]{%
        \includegraphics[width=0.24\linewidth]{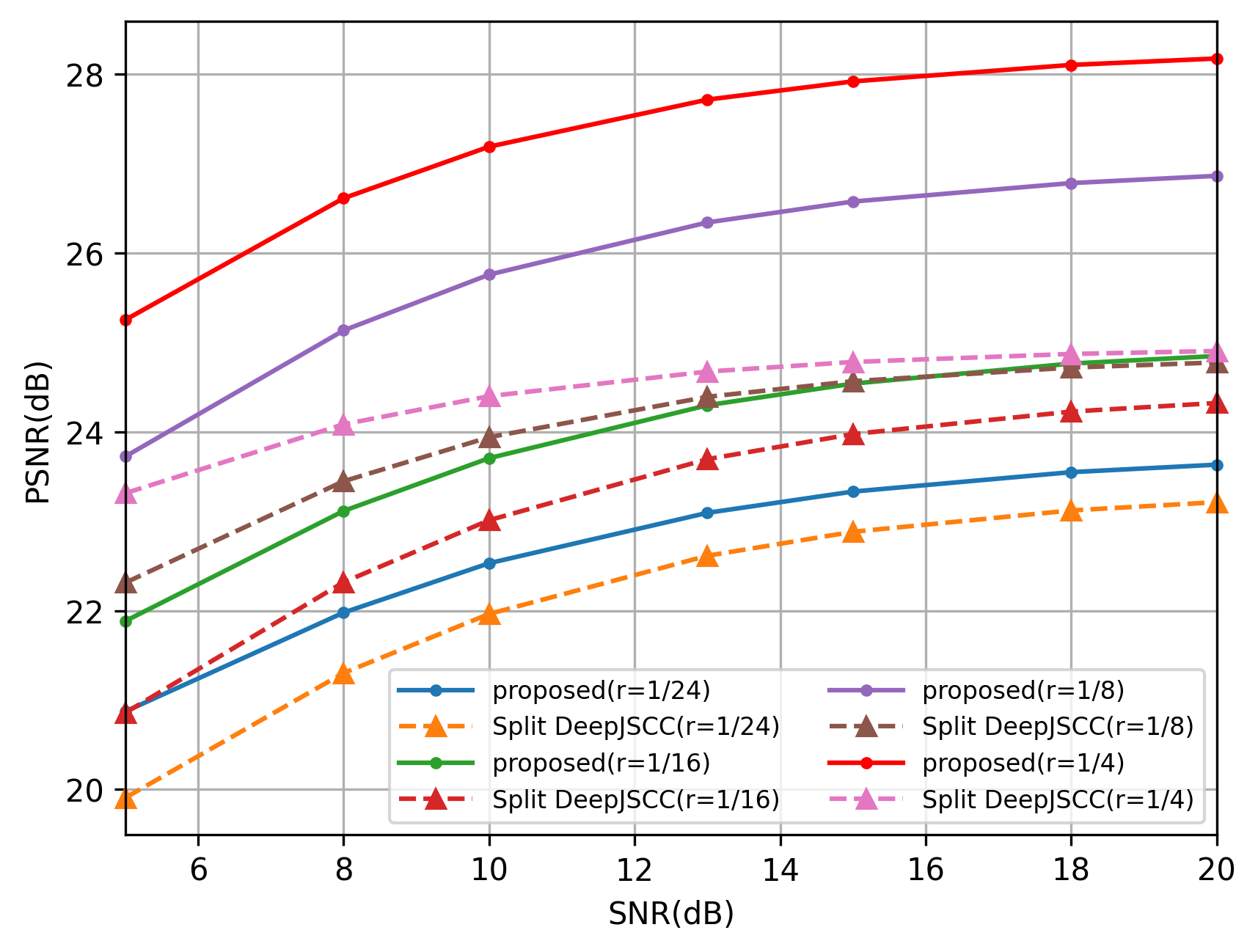}%
        \label{fig:AwgnImageNet32}%
    }
    \caption{PSNR versus SNR over AWGN channel}
    \label{fig:AWGN}
\end{figure*}

\begin{figure*}
    \centering
    \subfloat[CIFAR10]{%
        \includegraphics[width=0.24\linewidth]{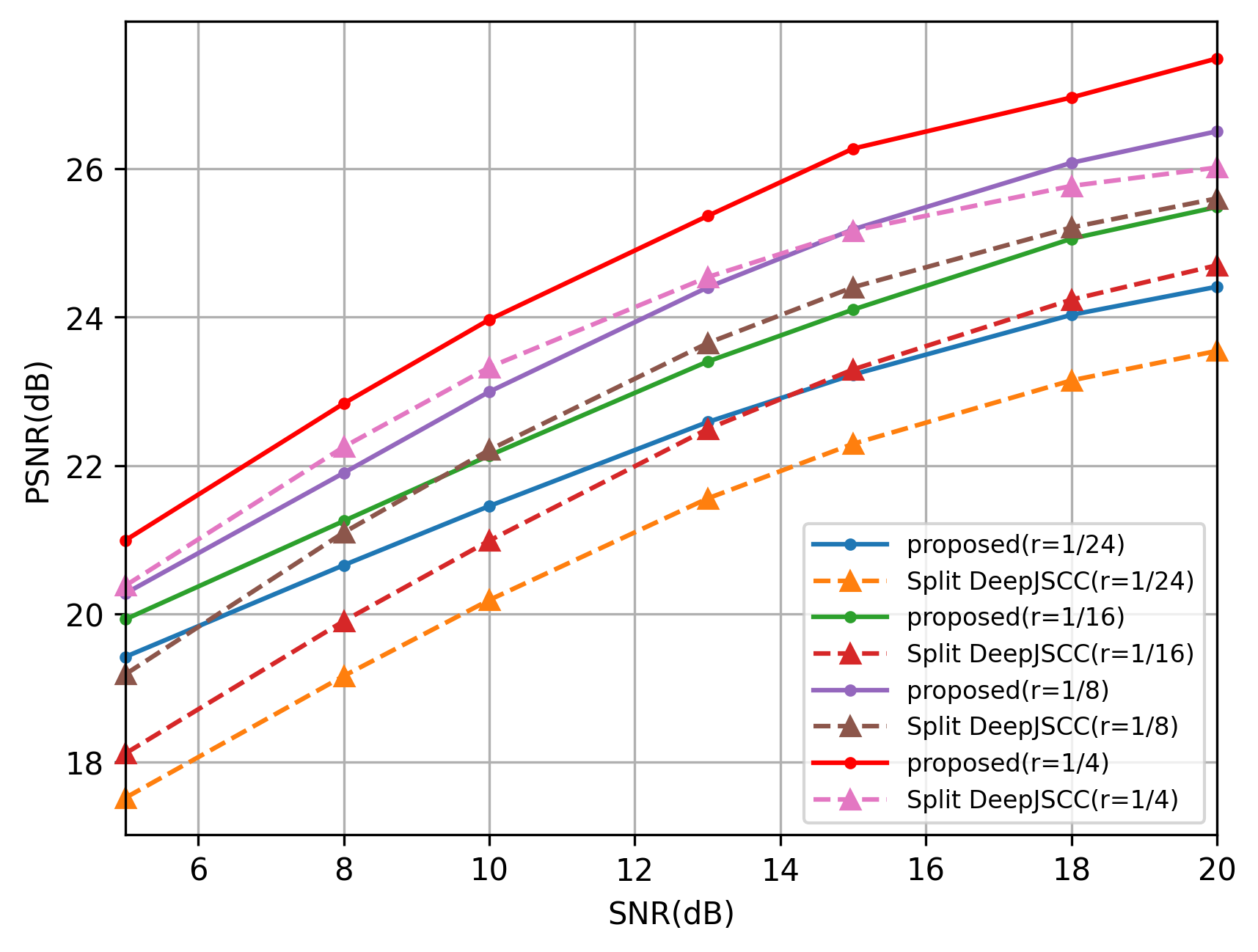}%
        \label{fig:RayleighCIFAR10}%
    }
    \hfil
    \subfloat[CIFAR100]{%
        \includegraphics[width=0.24\linewidth]{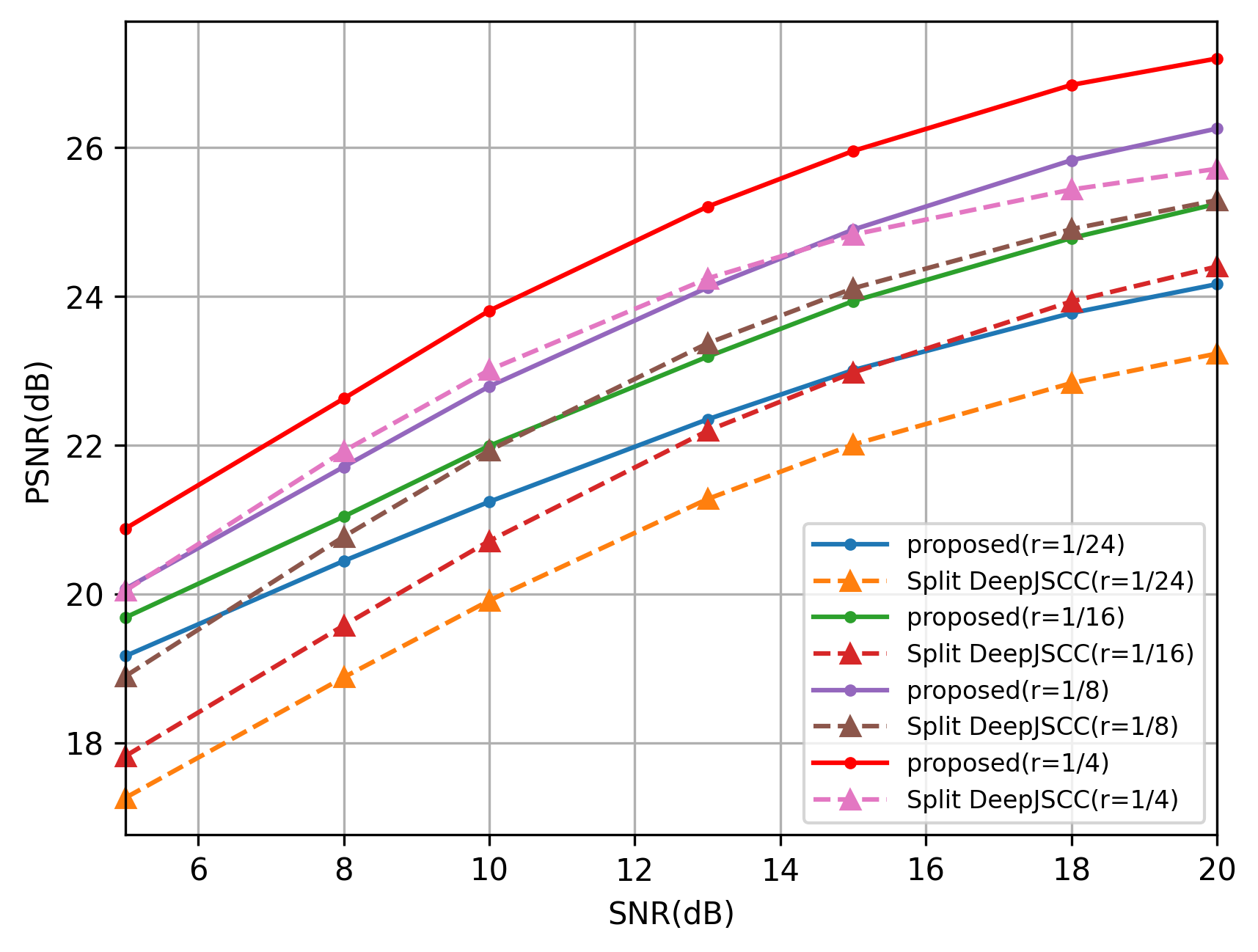}%
        \label{fig:RayleighCIFAR100}%
    }
    \hfil
    \subfloat[SVHN]{%
        \includegraphics[width=0.24\linewidth]{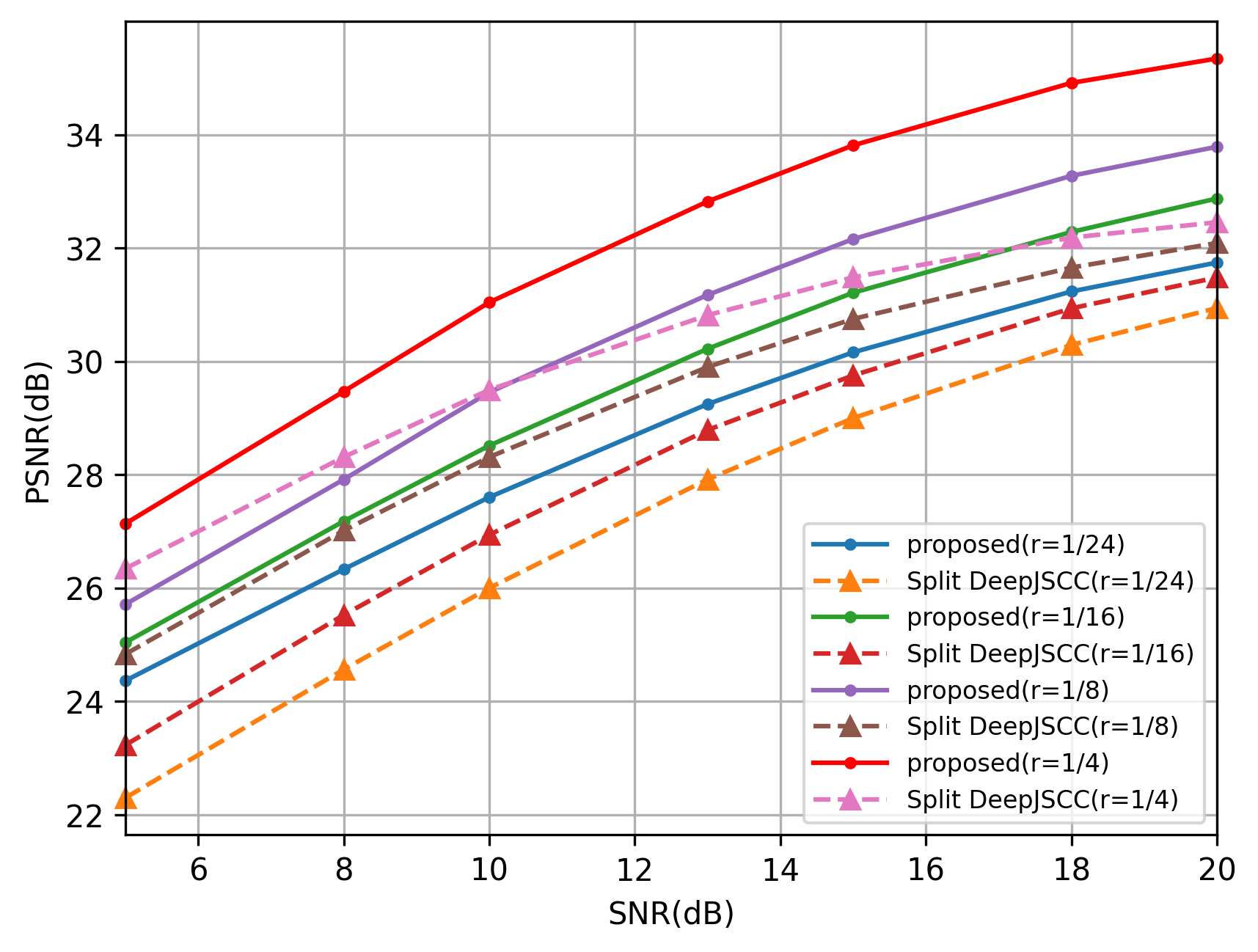}%
        \label{fig:RayleighSVHN}%
    }
    \hfil
    \subfloat[ImageNet32]{%
        \includegraphics[width=0.24\linewidth]{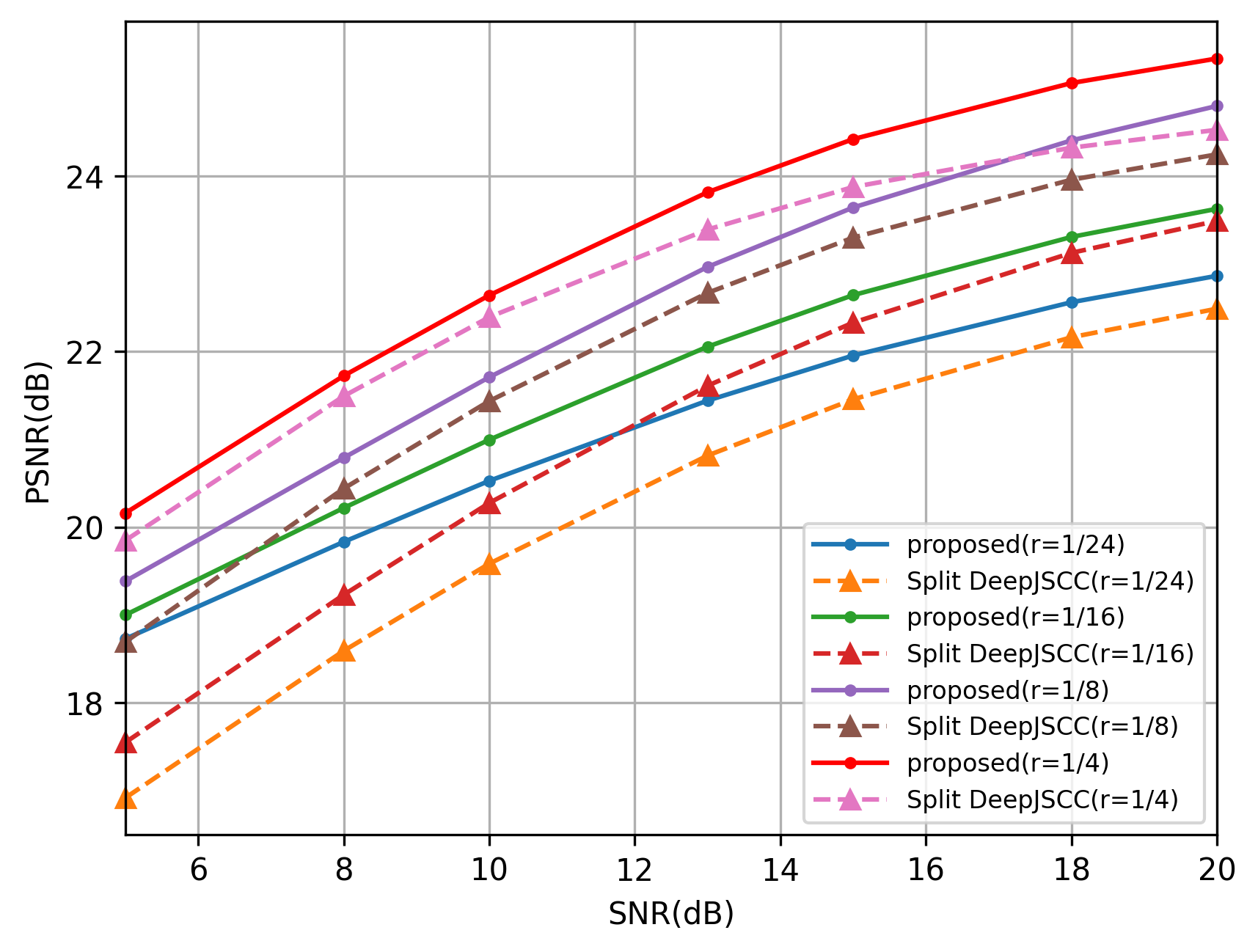}%
        \label{fig:RayleighImageNet32}%
    }
    \caption{PSNR versus SNR over Rayleigh fading channel}
    \label{fig:Rayleigh}
\end{figure*}
\subsubsection{Baselines}
We compare our proposed model’s performance with Split DeepJSCC \cite{tung2025multi}, which is also an interface-driven semantic image communication framework. The interface in Split DeepJSCC is prescribed without training, and the channel mapper and demapper modules do not leverage the importance information derived from the interface.
\subsubsection{Metrics}
We evaluate the performance of the proposed model using the widely used pixel-wise metric PSNR.
\subsubsection{Training Details}
In the first training stage, the model is trained for 250 epochs on CIFAR-10, 100 epochs on SVHN and ImageNet32. The regularization parameter $\lambda$ in \eqref{reg} is set to 1. In the second training stage, the number of epochs is set to 80 for all datasets. In both stages, we employ the Adam optimizer with a batch size of 128 and a learning rate of $1 \times 10^{-4}$ except for the Rayleigh fading channel, where the learning rate is adjusted to $5 \times 10^{-4}$. All experiments are conducted on an NVIDIA GeForce RTX 3090 GPU, and baseline methods (i.e., Split DeepJSCC) are trained and evaluated under identical configurations to ensure fair comparison.

For selection of training SNR, the proposed model is trained under the channel with a uniform distribution of SNR from 5dB to 20 dB. The baseline method (i.e., Split DeepJSCC) uses a fixed 20 dB SNR, which corresponds to their optimal performance setting.
\subsection{Performance Analysis}
In the evaluation, we test the proposed model under four CBRs (1/4, 1/8, 1/16, and 1/24) with SNR ranging from 5 dB to 20 dB across two channel conditions: AWGN and Rayleigh fading.

As shown in Fig.~\ref{fig:AWGN}, under AWGN channels, the PSNR performance of the proposed model outperforms Split DeepJSCC across all CBRs and SNRs on different datasets. While all the models exhibit performance improvements as CBR and SNR increase, our proposed model achieves larger gains, particularly under high CBR conditions. In particular, as the CBR increases, the performance gain of Split DeepJSCC becomes marginal. It can be observed that across different datasets, our proposed model exhibits performance improvements compared to Split DeepJSCC. Furthermore, although our proposed model is trained solely on the CIFAR-10 dataset, it performs well on both CIFAR-10 and CIFAR-100, demonstrating a certain level of generalization capability.

\begin{figure}[htbp]
    \centering
    \subfloat[AWGN]{%
        \includegraphics[width=0.24\textwidth]{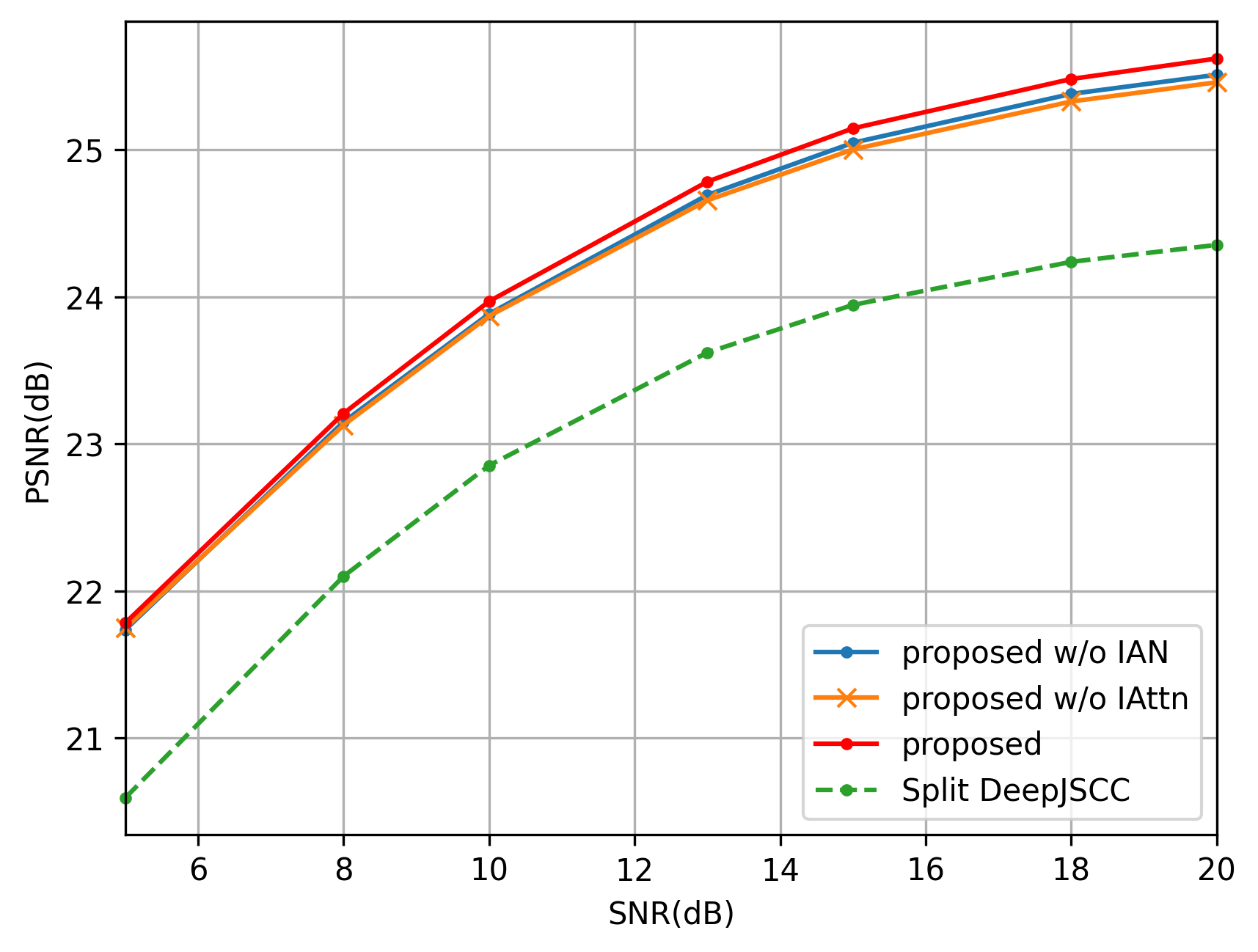}%
        \label{fig:ablation_awgn}%
    }
    \hfill
    \subfloat[Rayleigh fading]{%
        \includegraphics[width=0.24\textwidth]{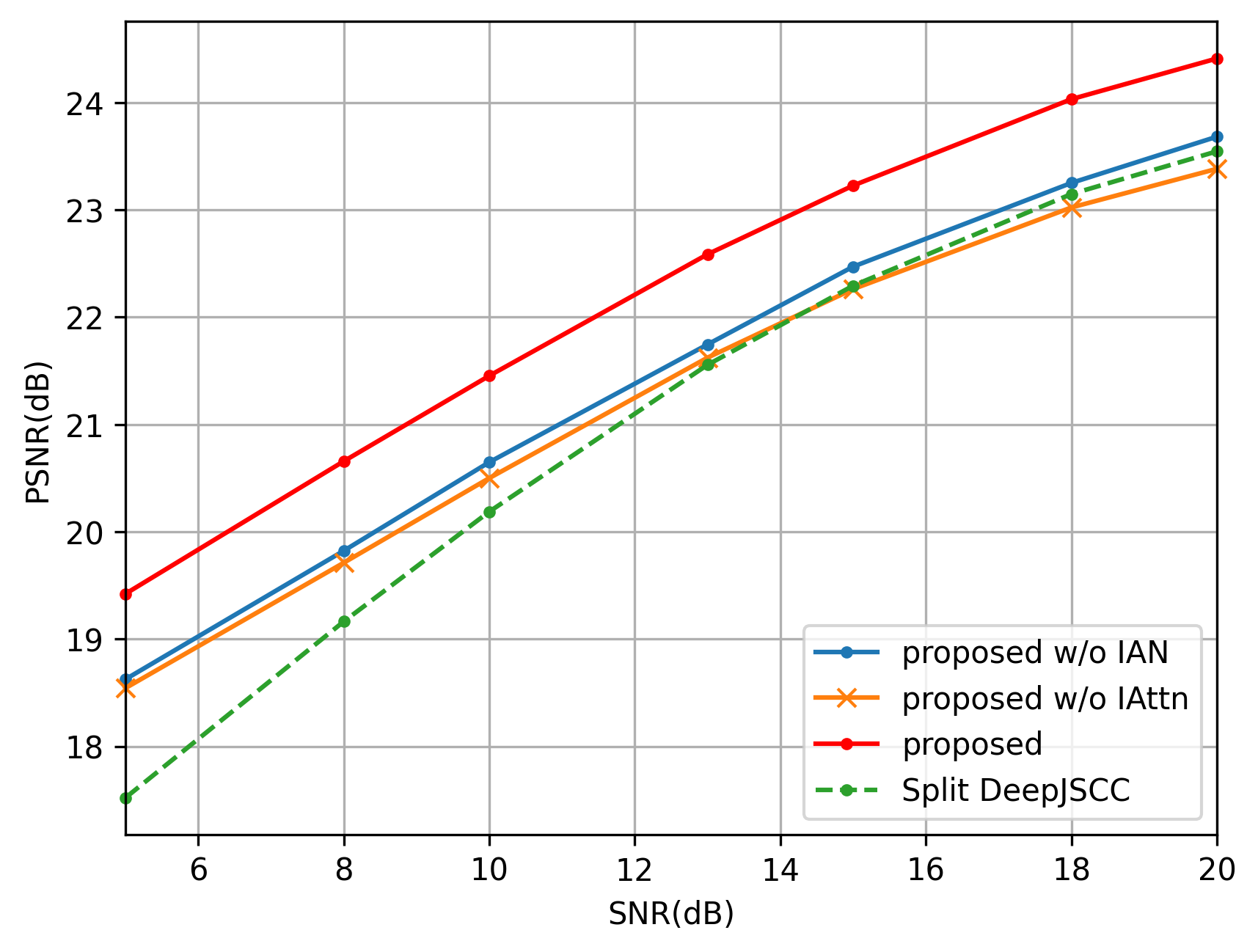}%
        \label{fig:ablation_rayleigh}%
    }
    \caption{PSNR comparison with/without Importance Aware Net.}
    \label{fig:ablation_IAN}
\end{figure}

Fig.~\ref{fig:Rayleigh} illustrates the results under Rayleigh fading channel. Consistent with the results observed in AWGN channels, our proposed model outperforms Split DeepJSCC in fading channel conditions. Notably, at lower CBRs, our porposed model exhibits slower performance degradation compared to Split DeepJSCC as SNR decreases. Moreover, at higher CBRs, our model achieves larger gains as SNR increases. This highlights the robustness of our proposed model in handling time-varying channel distortions. 

\begin{figure}[htbp]
    \centering
    \includegraphics[width=0.9\linewidth]{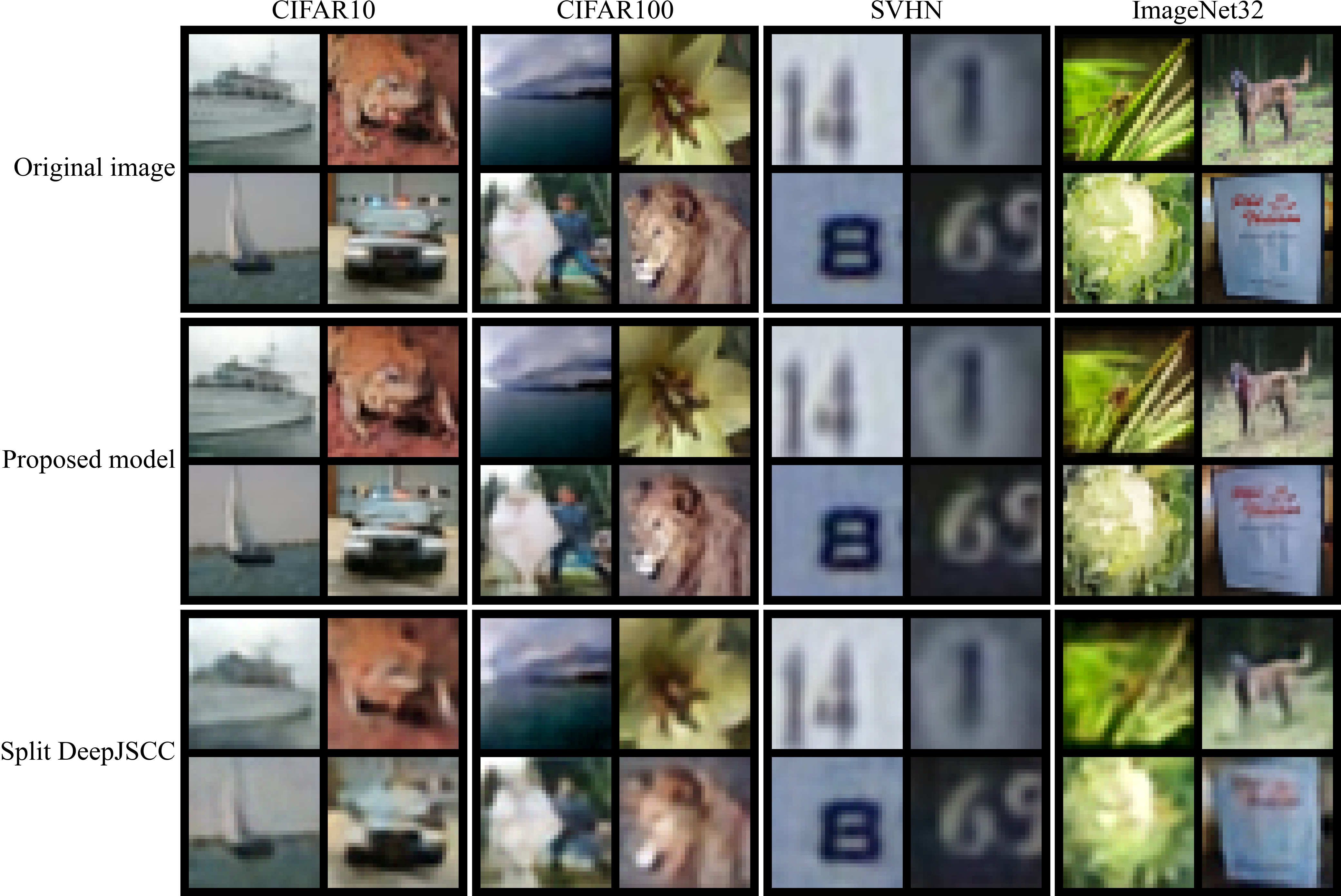}
    \caption{Examples of visual comparison.}
    \label{fig:visual}
\end{figure}

Fig.~\ref{fig:visual} shows the visual comparisons of reconstructed images across different datasets under AWGN channel at SNR = 20dB, with the CBR fixed at 1/4. It can be observed that our proposed model achieves better reconstruction quality on all the four datasets. Specifically, the reconstructed images retain more details and texture information while avoiding reconstruction artifacts such as blurring.

\subsection{Ablation Study}\label{ablation}
In order to show the effectiveness of our proposed IAN, we evaluated PSNR performance under two channel conditions (i.e., AWGN and Rayleigh fading) on the CIFAR10 dataset with the CBR fixed at 1/24 using four schemes: 1) \textbf{proposed:} our complete model; 2) \textbf{proposed w/o IAN:} the proposed model without the IAN (i.e., excluding both the Interface Attention Module and SE Block); 3) \textbf{proposed w/o IAttn:} the proposed model without Interface Attention Module; 4) \textbf{Split DeepJSCC:} the baseline Split DeepJSCC scheme. 

As shown in Fig~\ref{fig:ablation_IAN}, the proposed model exhibits consistent performance improvements across all SNR levels compared to proposed w/o IAN, proposed w/o IAttn and Split DeepJSCC. Moreover, removing either the Interface Attention Module or the IAN results in observable performance degradation, especially under Rayleigh fading channel. Specifically, the performance of the proposed model w/o IAttn is inferior to Split DeepJSCC at high SNR under the Rayleigh fading channel, and this strongly demonstrates the role of our proposed Interface Attention Module. Furthermore, the proposed model w/o IAttn (i.e., retaining only the SE Block) exhibits a performance degradation compared to the proposed model w/o IAN (i.e., removing both the Interface Attention Module and the SE Block). This further confirms that the best performance is achieved only when both the Interface Attention Module and the SE Block (i.e., the complete IAN) are used. This comparative analysis demonstrates that the our proposed IAN effectively enhances system robustness by capturing bit-level importance information provided by the learning-based interface, particularly under time-varying channel conditions. 

\section{Conclusion}
In this paper, we propose a learning-based interface for semantic communication, which effectively enhances the end-to-end performance of wireless image transmission compared with prescribed interface without training. Moreover, by introducing the IAN which utilizes the interface-derived bit importance information in the channel mapper and demapper, our approach achieves adaptive and efficient coding under varying CBRs and channel conditions. This work provides a potential solution for realizing semantic communications in existing wireless networks.

\bibliographystyle{IEEEtran}
\bibliography{IEEEabrv,ref}

\end{document}